\documentclass[showpacs,showkeys,11pt]{revtex4}
\usepackage{amsmath,amsthm,amssymb,amscd}
\usepackage{graphicx}
\newcommand{\al}{\alpha}

\newcommand{\la}{\lambda}
\newcommand{\om}{\omega}
\newcommand{\si}{\sigma}

\newcommand{\ze}{\zeta}
\newcommand{\Ga}{\Gamma}

\newcommand{\cJ}{{\mathcal J}}

\newcommand{\cY}{{\mathcal Y}}
\newcommand{\cW}{{\mathcal W}}

\def\ket#1{|#1\rangle}

\renewcommand{\ge}{\geqslant}

\renewcommand{\mod}{\operatorname{mod}}

\newcommand{\im}{\mathrm{Im}}

\newcounter{ex}

\begin{document}
\title{On the solutions of multicomponent generalizations \\ of the Lam\'{e} equation}
\author{J.C. \surname{Barba}}%
\email{jcbarba@fis.ucm.es} \affiliation{Departamento de F\'{\i}sica
  Te\'{o}rica II, Universidad Complutense, 28040 Madrid, Spain} \author{V.I. \surname{Inozemtsev}}%
\email[Corresponding author. Electronic address: ]{inozv@thsun1.jinr.ru}
 \affiliation{Laboratory of Theoretical Physics, JINR, 141980 Dubna, Moscow Region, Russia}
\date{\today}
\begin{abstract}
We describe a class of the singular solutions to the multicomponent analogs of
the Lam\'{e} equation, arising as equations of motion of the elliptic Calogero--Moser
systems of particles carrying spin $1/2$. At special value of the coupling
constant we propose the ansatz which allows one to get meromorphic solutions with
two arbitrary parameters. They are quantized upon the requirement of the
regularity of the wave function on the hyperplanes at which particles meet and
imposing periodic boundary conditions. We find also the extra integrals of motion
for three-particle systems which commute with the Hamiltonian for arbitrary
values of the coupling constant.
\end{abstract}
\maketitle
%
%
This letter is devoted to the problem of finding solutions to the matrix equation
which arises naturally in the theory of quantum Calogero--Moser $N$-particle
systems \cite{Ca75,Mo75},
\begin{equation}
H\psi = \Bigg[-\frac{1}{2}\sum_{j=1}^N \bigg( \frac{\partial}{\partial x_j}
\bigg)^2 + \sum_{j<k}^N a (a+S_{jk}) \wp(x_{jk}) \Bigg]\psi=E\psi,\label{E:H-N}
\end{equation}
where $\{x_j\}$ are coordinates of the particles, $x_{jk}\equiv x_j-x_k$, $\{s_j\}$ are their spins, $\psi$
depends on $\{x_j\}$ and $\{s_j\}$, $a\in\mathbf{R_+}$, $\{S_{jk}\}$ permute $s_j$ and $s_k$ and form the $SU(n)$ spin representation of the permutation group $S_N$, and $\wp(x)$ is the Weierstrass
elliptic function with two arbitrary periods $2\om_1,\,2\om_2$ with only
restriction $\im{(\om_2 \om_1^{-1})\ne 0}$. We shall suppose for definiteness
that $\om_1$ is real. At $N=2$, $\{S_{jk}\} \equiv 1$, one arrives at the usual
Lam\'{e} equation, but the matrix problem is highly nontrivial even in the case of
$N\ge 3,\,n=2$ (particles carry spin $1/2$) which we shall mainly discuss.

Various limits of the equation \eqref{E:H-N} were extensively studied in a lot of
papers
\cite{OP83,DI93,FV95,La06,Po92,HH92,Ka92,KW93,Ch94,HHTBP92,BGHP93,SS93,FGGRZ01}
and references therein. In \cite{OP83}, Olshanetsky and Perelomov found the
connection between \eqref{E:H-N} and the roots of $A_{N-1}$ algebra, and proved
the integrability of quantum spinless Calogero--Moser systems (i.e. $\{S_{jk}\}
\equiv 1$) but did not find any way of determining $\psi$. The first results for
spinless particles were obtained in \cite{DI93} where the explicit form of $\psi$
was found for $a=1,\, N=3$, and algebraic structure of the manifold containing
all $\psi$'s was described for all $a\in\bf{N}$, and all $N$. Later on,
overcomplicated meromorphic solution for all $a\in\bf{N}$, $N$ was found by
Felder and Varchenko \cite{FV95}. In \cite{La06}, the author obtained explicit
formulas for spinless case for all $a\in\bf{R_+}$ in the form of infinite series.

The trigonometric limit of \eqref{E:H-N} ($|\om_2|\to\infty$) has been also
intensively studied \cite{Po92,HH92,Ka92,KW93,Ch94,HHTBP92,BGHP93,SS93} for the
case of particles carrying spin. In \cite{Po92,HH92,Ka92,KW93,Ch94}, the
solutions were found as the spin generalization of the Jack polynomials which
provide adequate description in this limit of the eigenvalues problem for
spinless case and arbitrary $a\in\bf{R_+}$. The corresponding symmetry
responsible for these exact results was found to be the $\cY\big(SU(n)\big)$
Yangian algebra \cite{HHTBP92,BGHP93,BKW94}. The authors of the paper
\cite{FGGRZ01} have considered elliptic spin case but they have found the effect
of (quasi)exact solvability only for the case of the $BC_N$ root systems, and
\eqref{E:H-N} does not fall into their classes. The quantum Lax pair, i.e. the
solution to the equation $[H,L]=[L,M]$, $L,\,M$ being $(N\times N)$-matrices, was
also mentioned in \cite{SS93} for the elliptic case, but it was constructed
without any dependence on a spectral parameter. Moreover, in the elliptic case
even the existence of the Lax relation $[H,L]=[L,M]$ does not give the integrals
of motion in the form of $I_j=\sum_{k,l}^N(L^j)_{kl}$ since the $M$ matrix does
not obey the ``sum-to-zero" conditions $\sum_{j=1}^N M_{jk}=\sum_{k=1}^N
M_{jk}=0$.

In this situation, to our mind, every analytical results to the solutions of the elliptic matrix Schr\"{o}dinger equation \eqref{E:H-N} are of value, even for some restrictions for $a$, $N$ and $n$. In what follows, we put $N=3$ (three-particle case) and
consider at first the question of the integrability of the problem defined by
\eqref{E:H-N} for arbitrary $a$ and $n$. For the spinless case, it is known
\cite{OP83} that there is the operator
\begin{align}\label{E:JscalnoS}
\cJ_{scalar}=\frac{\partial^3}{\partial x_1 \partial x_2 \partial x_3} + a(a+1)
\bigg[ \big(  \wp(x_{23}) - \wp(\al) \big) \frac{\partial}{\partial x_1}&+ \big(
\wp(x_{31}) - \wp(\al) \big) \frac{\partial}{\partial x_2} \notag\\&+ \big(
\wp(x_{12}) - \wp(\al) \big) \frac{\partial}{\partial x_3}\bigg],
\end{align}
commuting with $H$ in this case, where $\al$ is an arbitrary ``spectral"
parameter which arises in the Lax-matrix approach. One can try to generalize this
structure for the spin case as
\begin{equation}
\widetilde\cJ_{scalar}=\frac{\partial^3}{\partial x_1 \partial x_2 \partial x_3}
+ \frac{1}{2}\sum_{j\ne k\ne l\ne j}^3 a(a+S_{jk})  \big(  \wp(x_{jk}) - \wp(\al)
\big) \frac{\partial}{\partial x_l},\label{E:JscalS}
\end{equation}
but direct calculation shows that \eqref{E:JscalS} does not commute with $H$ and
the term of higher order in permutations must be added. To gain some intuition
for constructing the proper operator, we consider the ``freezing trick''
($a\to\infty$) by which the integrals of motion for quantum elliptic spin chain
\cite{I9096} should be obtained, and try to add the analogs of these integrals to
the ansatz \eqref{E:JscalS},
\begin{align}
\widetilde\cJ_{spin}=\frac{\partial^3}{\partial x_1 \partial x_2 \partial x_3} &+
\frac{1}{2}\sum_{j\ne k\ne l\ne j}^3 a(a+S_{jk})  \big(  \wp(x_{jk}) - \wp(\al)
\big) \frac{\partial}{\partial x_l}  \notag\\ &+\la \sum_{j\ne k\ne l\ne j}^3
f(x_{jk}) f(x_{kl})f(x_{lj}) S_{jk} S_{kl},\label{E:Jspin}
\end{align}
where the function $f(x_{jk})$ is analogous to the elements of the Lax matrix for
quantum spin chain situation,
\begin{equation}
f(x_{jk})=\frac{\sigma(x_{jk}+\al)}{\si(x_{jk})\si(\al)}.\label{E:f_fun}
\end{equation}
Here $\si(x)$ is the Weierstrass sigma function,
$\frac{d^2}{dx^2}\log\si(x)=-\wp(x)$, $\si(x)\sim x+O(x^5)$ as $x\to 0$, and
$\la$ is some parameter which should be determined by the commutativity
condition,
\begin{equation}
[H,\widetilde\cJ_{spin}]=0.\notag
\end{equation}
Our computations shows that the ansatz \eqref{E:Jspin} is indeed correct and
\begin{equation}
\la=a^2/3.\label{E:lambda}
\end{equation}
Moreover, the function
\begin{equation}
\Phi_{jkl} =f(x_{jk}) f(x_{kl})f(x_{lj})\label{E:Phi_jkl1}
\end{equation}
is simplified drastically when being considered as the elliptic function of the
spectral parameter $\al$, having a pole of third order as $\al\to 0$ (and no
other singularities on the torus
$\mathbf{C}/\Ga,\,\Ga=2\mathbf{Z}\om_1+2\mathbf{Z}\om_2$). It can be written as
\begin{equation}
\Phi_{jkl}
=-\frac{1}{2}\wp'(\al)+\psi_{jkl}\wp(\al)+\varphi_{jkl},\label{E:Phi_jkl2}
\end{equation}
where the functions $\psi_{jkl}$ and $\varphi_{jkl}$ do not depend on $\al$ and
can be found by calculating the coefficients of the Laurent series for
\eqref{E:Phi_jkl1} at $\al\to 0$,
\begin{align}
\psi_{jkl} &= \ze(x_{jk})+\ze(x_{kl})+\ze(x_{lj}), \label{E:psi_jkl}\\
\varphi_{jkl} &= -\frac{1}{6} \Big\{\wp'(x_{jk}) + \wp'(x_{kl}) + \wp'(x_{lj}) +
2 \big[ \ze(x_{jk})+\ze(x_{kl})+\ze(x_{lj})\big]^3\Big\} \label{E:vphi_jkl}.
\end{align}
The formula \eqref{E:vphi_jkl} is obtained with the use of the well-known
relation
\begin{equation}
\wp(x_{jk}) + \wp(x_{kl}) + \wp(x_{lj}) =  \big[
\ze(x_{jk})+\ze(x_{kl})+\ze(x_{lj})\big]^2, \notag
\end{equation}
where $\ze(x)$ is the usual Weierstrass zeta function, $\ze'(x)=-\wp(x)$. Hence
we found  {\it two} independent integrals of motion from
(\ref{E:Jspin}--\ref{E:vphi_jkl}) due to the arbitrariness of a spectral
parameter $\al$,
\begin{align}
\cJ_{1}&=\sum_{j<k<l}^{3}\frac{\partial^3}{\partial x_j \partial x_k \partial x_l} +
\frac{1}{2}\sum_{j\ne k\ne l\ne j}^3 a(a+S_{jk}) \wp(x_{jk})
\frac{\partial}{\partial x_l} + \frac{a^2}{3}  \sum_{j\ne k\ne l\ne j}^3
\varphi_{jkl}S_{jk}S_{kl},\label{E:J1}\\
\cJ_{2}&=\frac{1}{2}\sum_{j\ne k\ne l\ne j}^3 S_{jk} \frac{\partial}{\partial
x_l} - \frac{a}{3} \sum_{j\ne k\ne l\ne j}^3
\psi_{jkl}S_{jk}S_{kl}.\label{E:J2}
\end{align}
The formula for $\cJ_{2}$ is especially simple: it resembles the total momentum
(and coincides with it as $\{S_{jk}\} = 1$). In the trigonometric limit, it can
be expressed through the scalar product of the Yangian generator and total spin
which (in this limit only!) both commute with $H$. We confirmed also, by direct
computation of $[\cJ_1,\cJ_2]$, that these operators mutually commute and form
with $H$ and total momentum the commutative ring for all values of the parameter
$a$.

Let us now construct the explicit solutions of \eqref{E:H-N} for the simplest
nontrivial case of three particles carrying spin $1/2$. When all spins aligned up
or down, we have the situation analogous to the spinless case \cite{DI93}. The
nontrivial form of the wave function $\psi$ arising for the states with total
spin $S=1/2$ is as follows
\begin{equation}
\psi(x_1,x_2,x_3)=A({\bf x})\ket{\uparrow\uparrow\downarrow}+B({\bf
x})\ket{\uparrow\downarrow\uparrow}+C({\bf
x})\ket{\downarrow\uparrow\uparrow},\quad A+B+C=0.\label{E:psi_x}
\end{equation}
The operators $\{a(a+S_{jk})\}$ act on the spin pairs in the states
$(\uparrow\uparrow),\,(\uparrow\downarrow+\downarrow\uparrow)$ as $a(a+1)$ and
for $(\uparrow\downarrow-\downarrow\uparrow)$ as $a(a-1)$. If $a$ is chosen as
positive integer, there are singularities in the spinless case in the form of
poles, $(x_j-x_k)^{-a}$ as $x_{jk}\to 0$. It is natural to expect that in the
case of particles with spin at least some solutions to \eqref{E:H-N} have the
similar structure, i.e. $A,\,B,\,C$ have singular behavior as $x_{jk}\to 0$ in
the form of poles. The equation \eqref{E:H-N} reads in the component form after
substituting \eqref{E:psi_x} as
\begin{align}
\Bigg(\frac{1}{2}\sum_{j=1}^3  \frac{\partial^2}{\partial x_j^2} + E \Bigg) A -
\sum_{j>k}^3 \wp(x_{jk}) A - \big[ \wp(x_{12}) A + \wp(x_{31}) C + \wp(x_{23}) B
\big] = 0, \label{E:A_difeq}\\
\Bigg(\frac{1}{2}\sum_{j=1}^3  \frac{\partial^2}{\partial x_j^2} + E \Bigg) B -
\sum_{j>k}^3 \wp(x_{jk}) B - \big[ \wp(x_{12}) C + \wp(x_{31}) B + \wp(x_{23}) A
\big] = 0, \label{E:B_difeq}\\
\Bigg(\frac{1}{2}\sum_{j=1}^3  \frac{\partial^2}{\partial x_j^2} + E \Bigg) C -
\sum_{j>k}^3 \wp(x_{jk}) C - \big[ \wp(x_{12}) B + \wp(x_{31}) A + \wp(x_{23}) C
\big] = 0.\label{E:C_difeq}
\end{align}
Let us introduce the notation
\begin{equation}
Y({\bf x})= A({\bf x})- B({\bf x}),\quad Z({\bf x})= A({\bf x})- C({\bf x})
\end{equation}
and deduct \eqref{E:B_difeq} and \eqref{E:C_difeq} from \eqref{E:A_difeq}. Under
the condition \eqref{E:psi_x}, it is easy to see that the system
(\ref{E:A_difeq}--\ref{E:C_difeq}) is equivalent to two coupled equations for
$Y({\bf x})$ and $Z({\bf x})$,
\begin{align}
\Bigg(\frac{1}{2}\sum_{j=1}^3  \frac{\partial^2}{\partial x_j^2} + E \Bigg) Y -
\wp(x_{12}) (Y+Z) -(2Y-Z) \wp(x_{31})  = 0, \label{E:Y_difeq}\\
\Bigg(\frac{1}{2}\sum_{j=1}^3  \frac{\partial^2}{\partial x_j^2} + E \Bigg) Z -
\wp(x_{12}) (Y+Z) -(2Z-Y) \wp(x_{23})  = 0. \label{E:Z_difeq}
\end{align}

Since the ``potentials'' here are double periodic, it might be expected that the
solutions to (\ref{E:Y_difeq},\ref{E:Z_difeq}) are quasiperiodic, acquiring the
same Bloch factors under the shifts of the arguments by the periods
$2\om_1,\,2\om_2$ of the Weierstrass functions. According to \eqref{E:Y_difeq},
$Y({\bf x})$ has a simple pole at $x_{31}\to 0$, the same is for $Z({\bf x})$ as
$x_{23}\to 0$. The analysis of limits $x_{23}\to 0$ for \eqref{E:Y_difeq} and
$x_{31}\to 0$ for \eqref{E:Z_difeq} shows that the left-hand sides of
\eqref{E:Y_difeq} and \eqref{E:Z_difeq} are regular at these conditions. And
finally, if $Y({\bf x}) \to Z({\bf x})$ as $x_{12}\to 0$, there should be a
simple pole singularity of these functions in this limit. Combining all these
properties, we come to the ansatz for $Y$ and $Z$ in the form
\begin{align}
Y({\bf x}) = b \frac{ \si(\mu_{12}) \si(x_{12}+\la_{12}) \si(x_{31}+\la_{31})} {\si(x_{12}) \si(x_{31})} \exp(k_1 x_1 + k_2 x_2 + k_3 x_3) , \label{E:Y_sol}\\
Z({\bf x}) = b \frac{ \si(\la_{12}) \si(x_{12}+\mu_{12}) \si(x_{23}+\mu_{23})}
{\si(x_{12}) \si(x_{23})} \exp(k_1 x_1 + k_2 x_2 + k_3 x_3),
\label{E:Z_sol}
\end{align}
where $\si(x)$ is the sigma Weierstrass function defined above,
$b,\,\{k_j\},\la_{12},\,\la_{31},\,\mu_{12},\,\mu_{23}$ are some parameters. The
Bloch factors for \eqref{E:Y_sol} and \eqref{E:Z_sol} are equal if and only if
\begin{equation}
\mu_{12}=\la_{12}-\la_{31},\quad\mu_{23}=-\la_{31}.\label{E:mu_cond}
\end{equation}
These expressions look rather asymmetric in $\{A,B,C\}$, but the symmetry becomes
evident with the use of the remarkable identity
\begin{align}
Y({\bf x}) - Z({\bf x}) &= C({\bf x}) - B({\bf x}) \notag\\&= - b \frac{
\si(\la_{31}) \si(x_{23}-\la_{12}) \si(x_{31}-\la_{12}+\la_{31})} {\si(x_{23})
\si(x_{31})} \exp(k_1 x_1 + k_2 x_2 + k_3 x_3) , \label{E:Y-Z_id}
\end{align}
which is valid for all values of the parameters $\la_{12}$ and $\la_{31}$ and
coordinates $\{x_{jk}\}$. Some long but not too tedious calculations show that
\eqref{E:Y_sol} and \eqref{E:Z_sol} under the condition \eqref{E:mu_cond} indeed
give the solutions to the system (\ref{E:Y_difeq},\ref{E:Z_difeq}) if the
following restrictions to the parameters $\{\la\},\,\{k\}$ take place,
\begin{align}
k_1-k_2&=\ze(\la_{31}-\la_{12})-\ze(\la_{12}),\label{E:k_cond1}\\
k_2-k_3&=\ze(\la_{31})+\ze(\la_{12}),\label{E:k_cond2}
\end{align}
where $\ze(x)$ is the zeta Weierstrass function defined above. To get (\ref{E:k_cond1},\ref{E:k_cond2}), we used the formula $$\ze(x)+\ze(y)+\ze(z)-\ze(x+y+z)= \frac{\si(x+y) \si(y+z) \si(z+x)}{\si(x) \si(y) \si(z) \si(x+y+z)}.$$ As for the
corresponding eigenvalue, it can be written in very symmetric form,
\begin{equation}
E=-\frac{1}{6}(k_1+k_2+k_3)^2-\frac{1}{3}\big( \wp(\la_{12}) + \wp(\la_{31}) +
\wp(\la_{12} - \la_{31}) \big)\label{E:energy},
\end{equation}
with two still unspecified parameters $\la_{12},\,\la_{31}$. Note that
(\ref{E:Y_sol},\ref{E:Z_sol},\ref{E:energy}) look very similar to the solution
of the usual Lam\'{e} equation,
\begin{equation}
-\frac{d^2 \psi(x)}{d x^2} + a(a+1) \psi(x) = E \psi(x),\notag
\end{equation}
in the case of $a=1$, where $\psi(x)\sim \exp\big(-x\ze(\al)\big) \si(x+\al)
\big[ \si(x) \big]^{-1}$ (the Hermite (1872) solution), and $E=-\wp(\al)$. We also found by some long
calculation that (\ref{E:Y_sol},\ref{E:Z_sol}) form also the eigenfunctions of
the operators $\cJ_1,\,\cJ_2$ (\ref{E:J1},\ref{E:J2}) at $a=1$, with the
eigenvalues
\begin{align}
j_1&=\frac{1}{27}(k_1+k_2+k_3)^3-\frac{k_1+k_2+k_3}{9}\big( \ze(\la_{12}) -
\ze(\la_{31}) - \ze(\la_{12} - \la_{31}) \big) \notag\\&\quad - \frac{1}{54}
\Big[ 14 \big(  \ze(\la_{12}) - \ze(\la_{31}) - \ze(\la_{12} - \la_{31}) \big)^3
+ 9 \big( \wp'(\la_{12}) - \wp'(\la_{31}) - \wp'(\la_{12} - \la_{31}) \big)
\Big],\label{E:j1_val}
\\ j_2&= \ze(\la_{12}) -
\ze(\la_{31}) - \ze(\la_{12} - \la_{31}),\label{E:j2_val}
\end{align}
also with arbitrary values of the parameters $\{\la\}$.

\begin{table}
\begin{center}
\begin{tabular}{@{\quad} c @{\qquad} c @{\quad} c @{\qquad} r @{\quad}}\hline\hline
 $(\tilde k_1,\tilde k_2,\tilde k_3)$ & $\tilde\la_{12}$ & $\tilde\la_{31}$ & $\psi({\bf x, s,\tilde k},\tilde\la_{12},\tilde\la_{31})$\\ \hline
 $(k_1,k_2,k_3)$ & $\la_{12}$ & $\la_{31}$ & $\psi({\bf x, s, k},\la_{12},\la_{31})$\\
 $(k_2,k_1,k_3)$ & $-(\la_{12}-\la_{31})$ & $\la_{31}$ & $-\Pi_{12}\psi({\bf x, s, k},\la_{12},\la_{31})$\\
 $(k_1,k_3,k_2)$ & $-\la_{31}$ & $-\la_{12}$ & $-\Pi_{23}\psi({\bf x, s, k},\la_{12},\la_{31})$\\
 $(k_3,k_2,k_1)$ & $\la_{12}$ & $\la_{12}-\la_{31}$ & $-\Pi_{31}\psi({\bf x, s, k},\la_{12},\la_{31})$ \\ $(k_2,k_3,k_1)$ &
 $-\la_{31}$ & $\la_{12}-\la_{31}$  & $\Pi_{12}\Pi_{23}\psi({\bf x, s, k},\la_{12},\la_{31})$ \\
 $(k_3,k_1,k_2)$ & $-(\la_{12}-\la_{31})$ & $-\la_{12}$ & $\Pi_{13}\Pi_{32}\psi({\bf x, s, k},\la_{12},\la_{31})$\\ \hline\hline
\end{tabular}
\end{center}\caption{Relation of the transformations of the parameters $\{k\}$ and $\{\la\}$ of the eigenfunction $\psi$ that obey (\ref{E:k_cond1}--\ref{E:energy})  and the action of permutation operators of particles $\{\Pi_{ij}\}$ on $\psi$.}\label{T:permut}\end{table}

Since the relation \eqref{E:energy} is invariant under all the permutations of the indices of $\{k\}$, the complete symmetrization of \eqref{E:psi_x} on permutations of particles also gives an eigenfunction of $H$, we named it $\psi_0({\bf x},{\bf s  },{\bf k},\la_{12},\la_{31})$, if $\{k\}$ and $\{\la\}$ obey (\ref{E:k_cond1}--\ref{E:energy}). The complete set of transformations of the parameters and its relation to permutation operators of particles is contained in Table \ref{T:permut}. This eigenfunction $\psi_0({\bf x},{\bf s},{\bf k},\la_{12},\la_{31})$ is regular as $x_j-x_k \to 0$. Furthermore, Eq. \eqref{E:energy} is invariant under global sign reversals of $\{k\}$ and $\{\la\}$ so we have another eigenfunction $\psi_1({\bf x},{\bf s},{\bf k},\la_{12},\la_{31})$ also symmetric under permutations of particles and regular as $x_j-x_k\to0$ and linearly independent of $\psi_0({\bf x},{\bf s},{\bf k},\la_{12},\la_{31})$, that verifies
\begin{equation}
 \psi_1({\bf x},{\bf s},{\bf k},\la_{12},\la_{31})=\psi_0({\bf x},{\bf s},-{\bf k},-\la_{12},-\la_{31})=-\psi_0(-{\bf x},{\bf s},{\bf k},\la_{12},\la_{31}).
\end{equation}

The discrete spectrum of the corresponding three-particle system on the real circle $x_{1,2,3}\in{\bf R}\mod2\om_1$ can be obtained by imposing the
periodic boundary conditions
\begin{equation}
\begin{aligned}(k_1-k_2)\om_1-\ze(\om_1)(\la_{12}-2\la_{31})&= {\rm i}\pi l_1\\(k_2-k_3)\om_1-\ze(\om_1)(\la_{12}+\la_{31})&= {\rm i}\pi l_2\end{aligned} \qquad l_1,l_2\in{\bf Z}.
\end{equation}

There is a degeneration in the energy, but $\psi_0$ and $\psi_1$ have distinct and opposite eigenvalues through the action of $\cJ_1$ and $\cJ_2$. Due to the relation satisfied by the total spin $$ \frac{1}{2} \otimes \frac{1}{2} \otimes \frac{1}{2} = \frac{3}{2} \oplus \frac{1}{2} \oplus \frac{1}{2},$$  there exists an intrinsic degeneration of functions with total spin $1/2$. That is the reason why we have such a degeneration on the energy level \eqref{E:energy}.

To conclude, we obtained for the first time the extra integrals of motion for the elliptic Calogero--Moser system of three particles with spin (\ref{E:J1},\ref{E:J2}). Thus we proved its complet integrability. It can be shown by direct computation with the use of the Liouville theorem that, replacing 3 by 4 in (\ref{E:J1},\ref{E:J2}), $\cJ_1$ and $\cJ_2$ are also two mutually commuting integrals of motion for the $N=4$ case. We found two non-trivial meromorphic eigenfunctions, depending on two parameters, for the spin $1/2$ case and coupling constant $a=1$ (\ref{E:Y_sol},\ref{E:Z_sol}).

We can conjecture that the regular eigenfunctions of \eqref{E:H-N} will be totally symmetric functions under permutation of particles for all values of $N$ and $a$ as it has been shown for some of its limits \cite{BGHP93} and for other unrelated spin dynamical models \cite{DG01}.
\medskip

\begin{acknowledgments}
This work was partially supported by the DGI under grant No.~FIS2005-00752, and
by the Complutense University and the DGUI under grant No.~GR74/07-910556. J.C.B.
acknowledges the financial support of the Spanish Ministry of Education and
Science through an FPU scholarship.

The work of V.I. was supported by the sabbatical grant of the Complutense University. V.I. would like also to thank Prof. Artemio Gonz\'{a}lez-L\'{o}pez for warm hospitality extended to him in Madrid where this work was done.

\vspace*{.5cm}
\end{acknowledgments}
\end{document}